\documentclass[aip,jcp,reprint,showkeys]{revtex4-1}
\usepackage{graphicx,dcolumn,bm,xcolor,microtype,hyperref,multirow,amscd,amsmath,amssymb,amsfonts,physics,mathpazo,mhchem}

\usepackage[normalem]{ulem}

\definecolor{darkgreen}{RGB}{0, 180, 0}

\newcommand{\cdash}{\multicolumn{1}{c}{---}}
\newcommand{\mc}{\multicolumn}
\newcommand{\mcc}[1]{\multicolumn{1}{c}{#1}}

\newcommand{\SI}{supplementary material}

\newcommand{\br}{\bm{r}}
\newcommand{\bR}{\bm{R}}

\newcommand{\EsCI}{E_\text{sCI}}

\newcommand{\EexFCI}{E_\text{exFCI}}

\newcommand{\sCI}[1]{sCI($#1$)}
\newcommand{\dT}{\text{Max}(\text{T})}
\newcommand{\dB}{\text{Max}(\text{B})}

\newcommand{\ex}[6]{$^{#1}#2_{#3}^{#4}(#5 \rightarrow #6)$}

\newcommand{\IneV}[1]{#1 eV}
\newcommand{\InAU}[1]{#1 a.u.}

\newcommand{\PsiExact}{\Phi}
\newcommand{\PsiT}{\Psi_\text{T}}

\newcommand{\bRUp}{\bR^{\uparrow}}
\newcommand{\bRDw}{\bR^{\downarrow}}
\newcommand{\Ndet}{N_\text{det}}
\newcommand{\NdetUp}{N_\text{det}^{\uparrow}}
\newcommand{\NdetDw}{N_\text{det}^{\downarrow}}

\newcommand{\sDetUp}[1]{D_{#1}^{\uparrow}}
\newcommand{\sDetDw}[1]{D_{#1}^{\downarrow}}

\newcommand{\NormUp}[1]{\mathcal{N}_{#1}^{\uparrow}}

\newcommand{\pis}{\pi^\star}

\newcommand{\LCPQ}{Laboratoire de Chimie et Physique Quantiques, Universit\'e de Toulouse, CNRS, UPS, France}
\newcommand{\CEISAM}{Laboratoire CEISAM - UMR CNRS 6230, Universit\'e de Nantes, 2 Rue de la Houssini\`ere, BP 92208, 44322 Nantes Cedex 3, France}
\newcommand{\CSD}{Computational Science Division, Argonne National Laboratory, Argonne, IL 60439, United States
of America}

\begin{document}	

\title{Excitation energies from diffusion Monte Carlo using selected Configuration Interaction nodes}	

\author{Anthony Scemama}
\affiliation{\LCPQ}
\author{Anouar Benali}
\affiliation{\CSD}
\author{Denis Jacquemin}
\affiliation{\CEISAM}    
\author{Michel Caffarel}
\affiliation{\LCPQ}
\author{Pierre-Fran{\c c}ois Loos}
\email[Corresponding author: ]{loos@irsamc.ups-tlse.fr}
\affiliation{\LCPQ}

\begin{abstract}
Quantum Monte Carlo (QMC) is a stochastic method which has been particularly successful for ground-state electronic structure calculations but mostly unexplored for the computation of excited-state energies.
Here, we show that, within a Jastrow-free QMC protocol relying on a deterministic and systematic construction of nodal surfaces using selected configuration interaction (sCI) expansions, one is able to obtain accurate excitation energies at the fixed-node diffusion Monte Carlo (FN-DMC) level.
This evidences that the fixed-node errors in the ground and excited states obtained with sCI wave functions cancel out to a large extent.
Our procedure is tested on two small organic molecules (water and formaldehyde) for which we report all-electron FN-DMC calculations.
For both the singlet and triplet manifolds, accurate vertical excitation energies are obtained with relatively compact multideterminant expansions built with small (typically double-$\zeta$) basis sets.
\end{abstract}

\keywords{diffusion Monte Carlo; fixed-node error; excited states; trial wave function}

\maketitle

%----------------------------------------------------------------
\section{Introduction}
%----------------------------------------------------------------
Processes related to electronically-excited states are central in chemistry, physics, and biology with applications, for example, in photochemistry, catalysis, solar cell technology, or light-driven biological processes (such as bioluminescence, photoisomerization, photosynthesis and others).

After mainly focussing on the calculation of ground-state energies and properties for half a century, accurate electronic structure theory methods have emerged for the computation of molecular excited states in the last decades.
There is no doubt that one of the main driving forces behind this evolution has been the emergence of the time-dependent version \cite{Casida} of density functional theory \cite{ParrBook} which has practically revolutionized computational quantum chemistry due to its user-friendly, black-box nature compared to more expensive complete active space methods (such as CASPT2 \cite{Roos, Andersson_1990}) where one has to choose an active space based on ``chemical intuition''.
However, fundamental deficiencies remain for the computation of extended conjugated systems, \cite{Woodcock_2002} charge-transfer states, \cite{Dreuw_2004} Rydberg states,\cite{Tozer_2000} doubly-excited states \cite{Tozer_2000} and others.
More expensive methods, such as CIS(D), \cite{Head-Gordon_1994, Head-Gordon_1995} CC2, \cite{Christiansen_1995, Hattig_2000} CC3, \cite{Christiansen_1995b, Koch_1997} ADC(2), \cite{Dreuw_2015} ADC(3), \cite{Harbach_2014} EOM-CCSD \cite{Purvis_1982} (and higher orders \cite{Noga_1987, Kucharski_1991}), have been designed to palliate these shortcomings, but they usually require large one-electron basis sets in order to provide converged results.	
Explicitly-correlated F12 versions of these methods requiring, by design, much smaller basis sets (but large auxiliary basis sets) are yet to become mainstream. \cite{Hattig12, Kong12} 

A method particularly successful for ground-state calculations but overlooked for excited states \cite{Grossman_2001, Porter_2001, Porter_2001a, Puzder_2002, Williamson_2002, Aspuru-Guzik_2004, Schautz_2004, Bande_2006, Bouabca_2009, Purwanto_2009, Zimmerman_2009, Dubecky_2010, Send_2011, Guareschi_2013, Dupuy_2015, Blunt_2017, Robinson_2017, Shea_2017, Zhao_2016, Scemama_2018} is quantum Monte Carlo (QMC), \cite{Kalos_1974, Ceperley_1979, Reynolds_1982, Foulkes_1999, Lester_2009, Austin_2012} and more particularly its diffusion version, DMC, based on the fixed-node (FN) approximation.
Within DMC, accurate calculations of vertical transition energies are tricky as one cannot rely on the variational principle in order to control the fixed-node error, contrary to methods such as configuration interaction (CI) for which one can safely lay on rigorous theorems such as the Hylleraas-Undheim and MacDonald theorem. \cite{Hylleraas_1930, McDonald_1933}
Moreover, the mechanism and degree of error compensation of the fixed-node error \cite{Ceperley_1991, Bressanini_2001, Rasch_2012, Rasch_2014, Kulahlioglu_2014} in the ground and excited states are mostly unknown, expect in a few cases. \cite{Bajdich_2005, Bressanini_2008, Scott_2007, Bressanini_2005a, Bressanini_2005b, Bressanini_2012, Mitas_2006, Scott_2007, Loos_2015}

Here, within our Jastrow-free QMC protocol relying on a deterministic construction of nodal surfaces, \cite{Giner_2015, Giner_2013, Caffarel_2016, Caffarel_2016b, Scemama_2018} we report all-electron FN-DMC calculations for the ground and excited states of water (\ce{H2O}) and formaldehyde (\ce{CH2O}) using large Dunning's basis sets including diffuse functions.
Our results for these two molecules evidence that one is able to obtain accurate excitation energies with relatively compact trial wave functions built with relatively small one-electron basis sets. 
Moreover, our approach has the advantage of being completely automatic and reproducible as one does not need to optimize the trial wave function \cite{Toulouse_2007, Toulouse_2008, Umrigar_2007} which is produced via a preliminary (deterministic) selected CI (sCI) method. \cite{Bender_1969, Whitten_1969, Huron_1973}
Recently, sCI methods have demonstrated their ability to reach near full CI (FCI) quality energies for small organic and transition metal-containing molecules. \cite{Giner_2013, Caffarel_2014, Giner_2015, Garniron_2017b, Caffarel_2016, Holmes_2016, Sharma_2017, Holmes_2017, Chien_2018, Scemama_2018, Loos_2018}

This manuscript is organized as follows.
Section \ref{sec:PsiT} provides details about the trial wave functions. 
Computational details are reported in Sec.~\ref{sec:details}.
In Section \ref{sec:res}, we discuss our results for water (Sec.~\ref{sec:water}) and formaldehyde (Sec.~\ref{sec:water}).
We draw our conclusion in Sec.~\ref{sec:ccl}.
Unless otherwise stated, atomic units are used.

%%% TABLE 0 %%%
\begin{table*}
	\caption{
		\label{tab:PsiT}
		Number of determinants $\Ndet$ (and their corresponding acronym) of the various sCI-based trial wave functions, denoted as \sCI{n}, for the singlet and triplet spin manifolds of \ce{H2O} and \ce{CH2O} at various truncation levels $\epsilon = 10^{-n}$.
		The size of the Hilbert space corresponding to the extrapolated FCI (exFCI) expansion is also reported. 
		exDMC is the extrapolated DMC energy obtained as described in the main text.
	}
		\begin{ruledtabular}
		\begin{tabular}{lcrrrrrrrrc}
			\mcc{Method}&	$\epsilon$		&	\mc{4}{c}{$\Ndet$ for singlet manifold}										&	\mc{4}{c}{$\Ndet$ for triplet manifold}																				&	acronym			\\
												\cline{3-6}																	\cline{7-10}
						&					&	\mc{3}{c}{\ce{H2O}}										&	\mc{1}{c}{\ce{CH2O}}				&	\mc{3}{c}{\ce{H2O}}									&	\mc{1}{c}{\ce{CH2O}}										\\
												\cline{3-5}													\cline{6-6}								\cline{7-9}											\cline{10-10}
						&					&	AVDZ			&	AVTZ			&	AVQZ			&	AVDZ			&	AVDZ			&	AVTZ			&	AVQZ		&	AVDZ										\\
			\hline

			sCI			&	$10^{-4}$		&	$9\,432$		&	$9\,948$		&	$8\,576$		&	$23\,317$		&	$5\,087$		&	$5\,760$		&	$5\,627$		&	$22\,938$		&	\sCI{4}			\\
						&	$10^{-5}$		&	$89\,797$		&	$110\,557$		&	$74\,414$		&	$255\,802$		&	$46\,264$		&	$58\,632$		&	$55\,637$		&	$227\,083$		&	\sCI{5}			\\
						&	$10^{-6}$		&	$636\,324$		&	$711\,120$		&	$325\,799$		&	$770\,978$		&	$234\,862$		&	$317\,880$		&	$243\,947$		&	$1\,074\,559$	&	\sCI{6}			\\
						&	$10^{-7}$		&	$3\,119\,643$	&	$2\,256\,057$	&	$697\,703$		&	\cdash 			&	$1\,029\,683$	&	$1\,074\,337$	&	$681\,392$		&	\cdash			&	\sCI{7}			\\
						&	$0$				&	$5\,869\,449$	&	$5\,589\,200$	&	$1\,139\,302$	&	$2\,043\,030$	&	$4\,566\,873$	&	$3\,760\,373$	&	$1\,833\,526$	&	$6\,637\,572$	&	sCI				\\
			exFCI		&	\textemdash		&	$\sim 10^{10}$	&	$\sim 10^{13}$	&	$\sim 10^{15}$	&	$\sim 10^{15}$	&	$\sim 10^{10}$	&	$\sim 10^{13}$	&	$\sim 10^{15}$	&	$\sim 10^{15}$	&	exDMC			\\
		\end{tabular}
		\end{ruledtabular}
\end{table*}
%%% %%% %%%

%----------------------------------------------------------------
\section{
Jastrow-free trial wave functions
\label{sec:PsiT}
}
%----------------------------------------------------------------
DMC is a stochastic projector technique, \cite{Kalos_1974, Ceperley_1979, Reynolds_1982} and its starting point is the time-dependent Schr\"odinger equation written in imaginary time $t$.
As $t \to \infty$, the steady state solution is the ground state wave function $\PsiExact(\bR)$ (where $\bR = (\br_1,\br_2,\ldots,\br_N)$ are the electron coordinates). \cite{Kolorenc_2011}
DMC generates configurations distributed according to the product $\PsiExact(\bR,t)\PsiT(\bR)$ and finds the best energy for the nodal surface of the trial wave function $\PsiT$. \cite{Loh_1990, Ceperley_1991, Troyer_2005} 
Therefore,  the ``nodal quality'' of $\PsiT$ is paramount in order to achieve high accuracy. 

Our trial wave functions have the particularity to be Jastrow-free, and they are simply written as a multideterminant expansion
\begin{equation}
	\label{eq:PsiT}
	\PsiT(\bR) = \sum_{i=1}^{\NdetUp} \sum_{j=1}^{\NdetDw} c_{ij} \sDetUp{i}(\bRUp) \sDetDw{j}(\bRDw),
\end{equation}
where the sets of spin-up and spin-down determinants, $\{D^{\uparrow}_i\}_{i=1,\ldots,\Ndet^\uparrow}$ and $\{D^{\downarrow}_i\}_{i=1,\ldots,\Ndet^\downarrow}$, and their corresponding coefficients $c_{ij}$ are generated via a preliminary sCI calculation. \cite{Huron_1973, Giner_2013, Giner_2015, Caffarel_2014, Scemama_2014, Caffarel_2016, Scemama_2016, Garniron_2017b, Scemama_2018} 
The absence of Jastrow factor eschews the non-linear stochastic optimization of $\PsiT(\bR)$. \cite{Toulouse_2007, Toulouse_2008, Umrigar_2007}
For a given set of molecular orbitals, the coefficients $c_{ij}$ originate from the diagonalization of a CI Hamiltonian matrix. 
Since the solution is unique (linear optimization), this defines a simple deterministic and systematic way of generating reproducible trial wave functions. 
Of course, if desired, a Jastow factor can be employed to reduce the statistical error as long as the multideterminant part is kept fixed (i.e.~the nodes are unchanged).

To avoid handling too many determinants in $\PsiT$, a truncation scheme is introduced which removes independently spin-up and spin-down determinants.
For example, a determinant $\sDetUp{i}$ is retained in $\PsiT$ if 
\begin{equation}
	\label{eq:trunc}
	\NormUp{i} = \sum_{j=1}^{\NdetDw} \abs{c_{ij}}^2 > \epsilon,
\end{equation}
where $\epsilon$ is a user-defined threshold. 
A similar formula is used for $\sDetDw{j}$.
When $\epsilon = 0$, the entire set of determinants is retained in the DMC simulation.
This truncation scheme is motivated by the fact that most of the computational effort lies in the calculation of the spin-specific determinants and their derivatives. Removing a product of determinants whose spin-specific determinants are already present in other products is insignificant regarding the computational cost. \cite{Scemama_2016}
For multi-state truncation, a natural generalization of the state-specific criterion defined by Eq.~\eqref{eq:trunc} is employed.
We refer the interested reader to Ref.~\onlinecite{Scemama_2018} for more details about trial wave function truncation.

The extrapolated DMC results, labeled as exDMC, have been obtained following our two-step extrapolation technique, as recently proposed in Ref.~\onlinecite{Scemama_2018}.
For a given basis set, the ultimate goal would be to use the nodes of the FCI wave function. 
This being out of reach, the sCI results are first extrapolated to the FCI limit using a second-order perturbation correction to the sCI energy. 
We refer to these extrapolated sCI results as exFCI.
Second, the DMC energies are obtained by performing a linear extrapolation as a function of $\EexFCI-\EsCI(n)$ for various values of the truncation
threshold $\epsilon=10^{-n}$. 
Additional details can be found in Refs.~\onlinecite{Scemama_2018, Loos_2018}.
All the total energies as well as the graphical representation of the various extrapolations performed in the present study can be found in the {\SI}.

%----------------------------------------------------------------
\section{
Computational details
\label{sec:details}
}
%----------------------------------------------------------------
The sCI calculations have been performed in the frozen-core approximation with the CIPSI (Configuration Interaction using a Perturbative Selection made Iteratively) algorithm \cite{Huron_1973} which uses a second-order perturbative criterion to select the energetically-relevant determinants in the FCI space. \cite{Giner_2013, Giner_2015, Caffarel_2014, Scemama_2014, Caffarel_2016, Scemama_2016, Garniron_2017b, Scemama_2018, Dash_2018}
In order to treat the electronic states of a given spin manifold on equal footing: i) all the singly-excited determinants are deterministically included at the start of the sCI calculation, and ii) a common set of determinants is used for the ground and excited states. \cite{Scemama_2018, Loos_2018}
This latter point is particularly important in practice.
An unbalanced treatment of the ground and excited states, even with different spatial symmetries, could have significant effects on the accuracy of the vertical transition energies. \cite{Loos_2018}
Moreover, to speed up convergence to the FCI limit, a common set of state-averaged natural orbitals issued from a preliminary (smaller) sCI calculation is employed.  

The geometries of \ce{H2O} and \ce{CH2O} have been obtained at the CC3/aug-cc-pVTZ level without frozen core approximation. 
These geometries have been extracted from Ref.~\onlinecite{Loos_2018}, and they are also reported in {\SI} for sake of completeness.
The augmented Dunning basis sets aug-cc-pVXZ (labeled as AVXZ in the following) with X $=$ D, T, and Q are used throughout this study.
The FN-DMC simulations are performed with the stochastic reconfiguration algorithm developed by Assaraf et al., \cite{Assaraf_2000} and a time-step of \InAU{$2 \times 10^{-4}$}
In the present case, it is not necessary to perform time-step extrapolations as the time-step error is smaller than the statistical error in the computation of excitation energies. 
To avoid further uncontrolled errors and technical difficulties introduced by pseudopotentials, we have performed all-electron DMC calculations.
A cusp correction is applied to all the molecular orbitals to ensure a well-behaved local energy at the electron-nucleus coalescence points, as described by Ma et al. \cite{Ma_2005} 

All sCI have been performed with the electronic structure software \textsc{quantum package}, \cite{QP} and the characteristics of the corresponding trial wave functions are gathered in Table \ref{tab:PsiT}. 
The QMC calculations have been performed with the \textsc{qmc=chem} suite of programs. \cite{qmcchem, Scemama_2013} 
Both softwares are developed in Toulouse and are freely available.

For both water and formaldehyde, the present FN-DMC results can be directly compared with the complete basis set (CBS) theoretical best estimates (TBEs) reported in Ref.~\onlinecite{Loos_2018}, which have been determined at the same CC3 geometry.
As explained in Ref.~\onlinecite{Loos_2018}, we believe that these TBE values have a typical error of the order of \IneV{0.03}, which is probably a generous upper bound in the case of compact molecules such as water and formaldehyde.
The experimental results --- extracted from Refs.~\onlinecite{Ralphs_2013} and \onlinecite{Robin_1985} for \ce{H2O} and \ce{CH2O} respectively --- only offer qualitative comparisons, for reasons discussed elsewhere. \cite{Dierksen_2004, Grimme_2004, Santoro_2016, Ralphs_2013, Loos_2018}
%----------------------------------------------------------------
\section{
Results and discussion
\label{sec:res}
}
%----------------------------------------------------------------
%----------------------------------------------------------------
\subsection{
Water
\label{sec:water}
}
%----------------------------------------------------------------

%%% TABLE I %%%
\begin{table}
\caption{
\label{tab:H2O-DMC}
Vertical excitation energies (in eV) for the three lowest singlet and three lowest triplet excited states of water obtained with the all-electron AVXZ basis sets (X $=$ D, T and Q) for various trial wave functions $\PsiT$ (see Table \ref{tab:PsiT}).
The error bar corresponding to one standard error is reported in parenthesis.
For a given transition, $\dB$ and $\dT$ are the maximum absolute deviation between excitation energies for a given basis and for a given trial wave function, respectively.
} 
\begin{ruledtabular}
\begin{tabular}{lcrrrc}
Transition				&	$\PsiT$		& \mc{3}{c}{Dunning's basis set}				&	$\dT$		\\
										\cline{3-5}
						&				&	AVDZ 		&	AVTZ		&	AVQZ		&			\\							
\hline
\ex{1}{B}{1}{}{n}{3s}	&	\sCI{4}		&	$7.70(1)$	&	$7.69(1)$	&	$7.70(1)$	&	$0.01(1)$	\\
 						&	\sCI{5}		&	$7.73(1)$	&	$7.72(1)$	&	$7.74(1)$	&	$0.02(1)$	\\
						&	\sCI{6}		&	$7.73(1)$	&	$7.69(2)$	&	$7.71(1)$	&	$0.04(2)$	\\
						&	\sCI{7}		&	$7.69(3)$	&	$7.71(2)$	&	$7.73(1)$	&	$0.04(3)$	\\
						&	exDMC		&	$7.73(1)$	&	$7.70(2)$	&	$7.71(1)$	&	$0.03(2)$	\\
	$\dB$				&				&	\mcc{$0.04(3)$}&	\mcc{$0.03(2)$}&	\mcc{$0.04(1)$}&	\\
						\hline
\ex{1}{A}{2}{}{n}{3p}	&	\sCI{4}		&	$9.50(1)$	&	$9.49(1)$	&	$9.48(1)$	&	$0.02(1)$	\\
						&	\sCI{5}		&	$9.48(1)$	&	$9.51(1)$	&	$9.51(1)$	&	$0.03(1)$	\\
						&	\sCI{6}		&	$9.49(1)$	&	$9.50(2)$	&	$9.47(1)$	&	$0.03(2)$	\\
						&	\sCI{7}		&	$9.43(3)$	&	$9.46(2)$	&	$9.49(1)$	&	$0.06(3)$	\\
						&	exDMC		&	$9.48(1)$	&	$9.47(2)$	&	$9.47(1)$	&	$0.01(2)$	\\
	$\dB$				&				&	\mcc{$0.07(3)$}&	\mcc{$0.06(2)$}&	\mcc{$0.04(1)$}&	\\
						\hline
\ex{1}{A}{1}{}{n}{3s}	&	\sCI{4}		&	$10.11(1)$	&	$10.09(1)$	&	$10.08(1)$	&	$0.03(1)$	\\
						&	\sCI{5}		&	$10.09(1)$	&	$10.10(1)$	&	$10.10(1)$	&	$0.01(1)$	\\
						&	\sCI{6}		&	$10.10(1)$	&	$10.06(2)$	&	$10.05(1)$	&	$0.05(1)$	\\
						&	\sCI{7}		&	$10.09(3)$	&	$10.07(2)$	&	$10.06(1)$	&	$0.04(3)$	\\
						&	exDMC		&	$10.10(1)$	&	$10.05(2)$	&	$10.03(1)$	&	$0.07(2)$	\\
	$\dB$				&				&	\mcc{$0.02(3)$}&	\mcc{$0.05(2)$}&	\mcc{$0.07(1)$}&	\\
						\hline
\ex{3}{B}{1}{}{n}{3s}	&	\sCI{4}		&	$7.31(1)$	&	$7.31(1)$	&	$7.30(1)$	&	$0.01(1)$	\\
						&	\sCI{5}		&	$7.33(1)$	&	$7.33(1)$	&	$7.31(1)$	&	$0.02(1)$	\\
						&	\sCI{6}		&	$7.35(1)$	&	$7.31(1)$	&	$7.29(1)$	&	$0.06(1)$	\\
						&	\sCI{7}		&	$7.35(2)$	&	$7.37(2)$	&	$7.30(1)$	&	$0.07(2)$	\\
						&	exDMC		&	$7.36(1)$	&	$7.35(1)$	&	$7.30(1)$	&	$0.06(1)$	\\
	$\dB$				&				&	\mcc{$0.05(1)$}&	\mcc{$0.06(2)$}&	\mcc{$0.02(1)$}&	\\
						\hline
\ex{3}{A}{2}{}{n}{3p}	&	\sCI{4}		&	$9.27(1)$	&	$9.28(1)$	&	$9.28(1)$	&	$0.01(1)$	\\
						&	\sCI{5}		&	$9.31(1)$	&	$9.32(1)$	&	$9.31(1)$	&	$0.01(1)$	\\
						&	\sCI{6}		&	$9.33(1)$	&	$9.31(1)$	&	$9.28(1)$	&	$0.05(1)$	\\
						&	\sCI{7}		&	$9.30(2)$	&	$9.34(2)$	&	$9.28(1)$	&	$0.06(2)$	\\
						&	exDMC		&	$9.33(1)$	&	$9.32(1)$	&	$9.28(1)$	&	$0.05(2)$	\\
	$\dB$				&				&	\mcc{$0.06(1)$}&	\mcc{$0.06(2)$}&	\mcc{$0.03(1)$}&	\\
						\hline
\ex{3}{A}{1}{}{n}{3s}	&	\sCI{4}		&	$9.59(1)$	&	$9.58(1)$	&	$9.58(1)$	&	$0.01(1)$	\\
						&	\sCI{5}		&	$9.61(1)$	&	$9.60(1)$	&	$9.58(1)$	&	$0.03(1)$	\\
						&	\sCI{6}		&	$9.62(1)$	&	$9.59(1)$	&	$9.58(1)$	&	$0.04(1)$	\\
						&	\sCI{7}		&	$9.61(3)$	&	$9.60(2)$	&	$9.57(1)$	&	$0.04(3)$	\\
						&	exDMC		&	$9.63(1)$	&	$9.60(1)$	&	$9.59(1)$	&	$0.04(2)$	\\
	$\dB$				&				&	\mcc{$0.04(1)$}&	\mcc{$0.02(1)$}&	\mcc{$0.02(1)$}&	\\
\end{tabular}
\end{ruledtabular}
\end{table}
%%% %%% %%%

%%% TABLE II %%%
\begin{table*}
\caption{
\label{tab:H2O-Extrap}
Extrapolated vertical excitation energies (in eV) for the three lowest singlet and three lowest triplet excited states of water obtained with the all-electron AVXZ basis sets (X $=$ D, T and Q).
The error bar corresponding to one standard error is reported in parenthesis.
} 
\begin{ruledtabular}
\begin{tabular}{lrrrrrrrrcc}
Transition				& \multicolumn{2}{c}{AVDZ} 	& \multicolumn{2}{c}{AVTZ}	& \multicolumn{2}{c}{AVQZ}	& \multicolumn{2}{c}{CBS}	
						&	TBE\footnotemark[1]		&	Exp.\footnotemark[2]	\\
						\cline{2-3}					\cline{4-5}					\cline{6-7}					\cline{8-9}		
							
	 					&	exFCI 	&	exDMC 		&	exFCI 	&	exDMC 		&	exFCI 	&	exDMC  		&	exFCI 	&	exDMC  		&			&	\\
\hline
\ex{1}{B}{1}{}{n}{3s} 	&	$7.53$	&	$7.73(1)$	&	$7.63$	&	$7.70(2)$	&	$7.68$	&	$7.71(1)$	&	$7.70$	&	$7.70(1)$	&	$7.70$	&	$7.41$	\\
\ex{1}{A}{2}{}{n}{3p} 	&	$9.32$	&	$9.48(1)$	&	$9.41$	&	$9.47(2)$	&	$9.46$	&	$9.47(1)$	&	$9.48$	&	$9.46(1)$	&	$9.47$	&	$9.20$	\\
\ex{1}{A}{1}{}{n}{3s} 	&	$9.94$	&	$10.10(1)$	&	$9.99$	&	$10.05(2)$	&	$10.03$	&	$10.03(1)$	&	$10.03$	&	$10.01(1)$	&	$9.97$	&	$9.67$	\\
\hline
\ex{3}{B}{1}{}{n}{3s} 	&	$7.14$	&	$7.36(1)$	&	$7.25$	&	$7.35(1)$	&	$7.30$	&	$7.30(1)$	&	$7.31$	&	$7.30(1)$	&	$7.33$	&	$7.20$	\\
\ex{3}{A}{2}{}{n}{3p} 	&	$9.14$	&	$9.33(1)$	&	$9.24$	&	$9.32(1)$	&	$9.29$	&	$9.28(1)$	&	$9.30$	&	$9.28(1)$	&	$9.30$	&	$8.90$	\\
\ex{3}{A}{1}{}{n}{3s} 	&	$9.48$	&	$9.63(1)$	&	$9.54$	&	$9.61(1)$	&	$9.58$	&	$9.59(1)$	&	$9.58$	&	$9.57(1)$	&	$9.59$	&	$9.46$	\\
\end{tabular}
\end{ruledtabular}
\footnotetext[2]{Theoretical best estimates of Ref.~\onlinecite{Loos_2018} obtained from exFCI/AVQZ data corrected with the difference between CC3/AVQZ and CC3/d-aug-cc-pV5Z values.}
\footnotetext[1]{Energy loss experiment from Ref.~\citenum{Ralphs_2013}.}
\end{table*}
%%% %%% %%%

%%% FIG 1 %%
\begin{figure*}
	\includegraphics[width=0.8\linewidth]{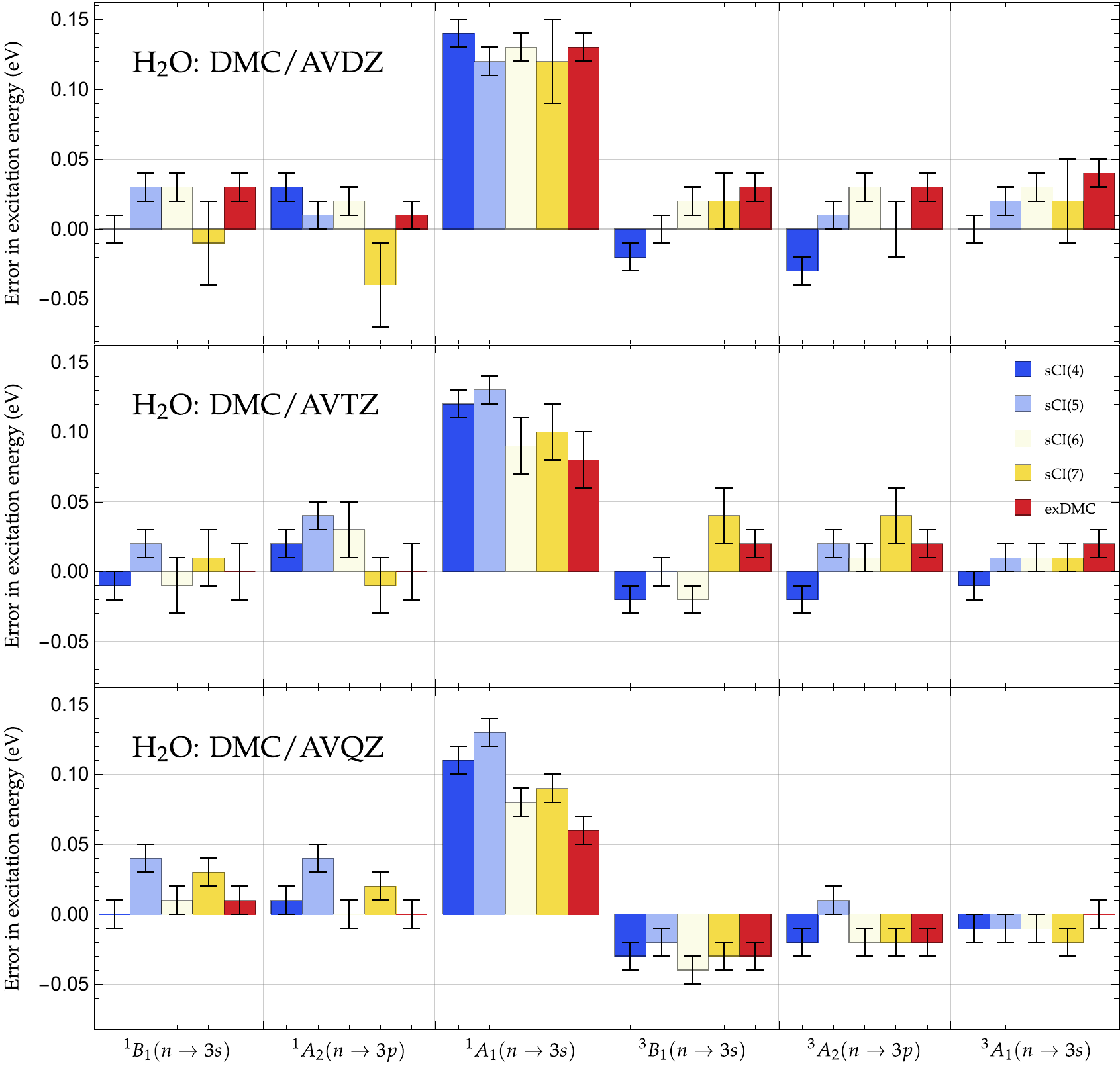}
	\caption{
	\label{fig:FigH2O_1}
	Error (in eV) compared to the TBE of Ref.~\onlinecite{Loos_2018} for the three lowest singlet and three lowest triplet DMC excitation energies of the water molecule computed with the AVDZ (top), AVTZ (center) and AVQZ (bottom) basis sets and various trial wave functions (see Table \ref{tab:PsiT}). 
	The error bar corresponds to one standard error.
	}
\end{figure*}
%%% %%% %%%

%%% FIG 2 %%
\begin{figure*}
	\includegraphics[width=0.8\linewidth]{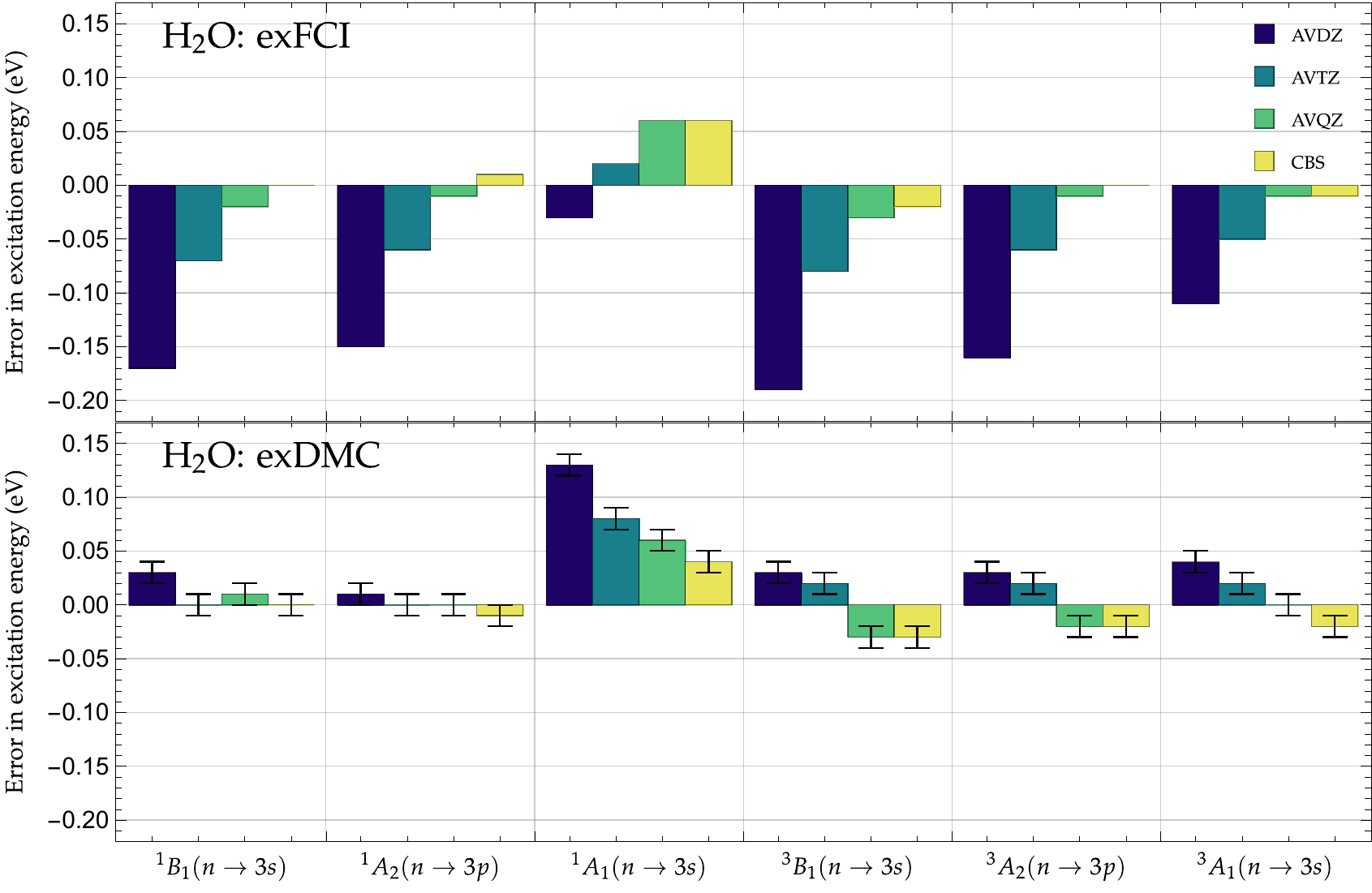}
	\caption{
	\label{fig:FigH2O_2}
	Error (in eV) compared to the TBE of Ref.~\onlinecite{Loos_2018} for the three lowest singlet and three lowest triplet exFCI (top) and exDMC (bottom) vertical excitation energies of the water molecule computed with various basis sets.
	The CBS values (in yellow) are also reported. 
	The error bar corresponds to one standard error.
	}
\end{figure*}
%%% %%% %%%

The excited states of the water molecule are often used as a test case for Rydberg excitations and they have been thoroughly studied at various levels of theory in the past two decades. \cite{Leang_2012, Hoyer_2016, Cai_2000c, Li_2006b, Rubio_2008, Palenikova_2008, Loos_2018}  
We have computed the vertical transition energy of the first three lowest singlet and three lowest triplet states of \ce{H2O}.
All these excited states have a Rydberg character, hence exhibiting a strong basis set sensitivity.

In Table \ref{tab:H2O-DMC}, we report all-electron FN-DMC estimates of the vertical excitation energies for various trial wave functions ({\sCI{n}} with $n=4$--$7$) and various diffuse Dunning's basis sets (AVXZ with X $=$ D, T and Q).
The error in the excitation energies compared to the TBEs reported in our recent benchmark study \cite{Loos_2018} are depicted in Fig.~\ref{fig:FigH2O_1}.	
Table \ref{tab:H2O-DMC} also reports the extrapolated DMC (exDMC) excitation energies.
As mentioned above, we refer the interested reader to the {\SI} for additional details about the extrapolation procedure (see also Ref.~\onlinecite{Scemama_2018}). 

Table \ref{tab:H2O-DMC} also reports, for each transition, the maximum absolute deviation between excitation energies for a given basis [$\dB$] and for a given trial wave function [$\dT$].
Note that, although the total energies do change significantly, the values of the excitation energies are fairly stable with respect to the number of determinants in $\PsiT$ (see Fig.~\ref{fig:FigH2O_1}).
Indeed, the maximum value of $\dT$ is only \IneV{$0.07$}, and can be as small as \IneV{$0.02$} in certain cases.
This shows that, even for small trial wave functions, we have a large amount of error cancelation in the fixed-node error of the ground and excited states.
Similarly, at the DMC level, the excitation energies are weakly basis dependent, as evidenced by the stability of the results with respect to the one-electron basis.
For a given state and trial wave function, the maximum value of $\dB$ is \IneV{$0.07$}, falling down to \IneV{$0.01$} in certain cases.

In Fig.~\ref{fig:FigH2O_2}, we report the basis set convergence of our extrapolated DMC results (exDMC) as well as the extrapolated sCI calculations (exFCI).
We have also reported the CBS results obtained by the usual $(X+1/2)^{-3}$ extrapolation. \cite{HelgakerBook}
The graphs associated with these CBS extrapolations can be found in the {\SI}, where one can directly notice the quality of these fits.

One would have noted that the results for the \ex{1}{A}{1}{}{n}{3p} transition are significantly worse than the others.
%It is well known that totally-symmetric excited states are very basis set dependent.
This can be explained by a particularly strong basis set effect.
Indeed, we have recently shown that, even within conventional deterministic wave function methods such as high-level coupled cluster theories, this particular state requires doubly-augmented basis sets (d-aug-cc-pVXZ) in order to describe properly its strong Rydberg character. \cite{Loos_2018}
Therefore, at the DMC/AVDZ level of theory, we have an error greater than \IneV{$0.1$}.
However, when one improves the basis description, DMC is going in the right direction.
In the CBS limit, the error drops below \IneV{$0.05$}, even without doubly-augmented basis sets.

For all the other excitations, we have a maximum error of \IneV{$0.04$} compared to the TBE, and this drops to \IneV{$0.03$} in the CBS limit (see Fig.~\ref{fig:FigH2O_2}).
From a practical point of view, it is important to emphasize that it is generally possible to obtain accurate vertical transition energies with the compact AVDZ basis set and relatively small multideterminant expansions.
In other words, for this particular molecule, chemical accuracy (error below $1$ kcal/mol) is generally attainable within FN-DMC with a small (i.e. double-$\zeta$)  basis set.

%----------------------------------------------------------------
\subsection{
Formaldehyde
\label{sec:formaldehyde}
}
%----------------------------------------------------------------

%%% TABLE VI %%%
\begin{table*}
\caption{
\label{tab:CH2O-DMC}
Vertical excitation energies (in eV) and their correspoding nature (valence or Rydberg) for the five lowest singlet and five lowest triplet excited states of formaldehyde obtained with the all-electron AVDZ basis set.
The error with respect to the TBE reported in Ref.~\onlinecite{Loos_2018} is also reported for the exFCI and exDMC calculations.
The error bar corresponding to one standard error is reported in parenthesis.
} 
\begin{ruledtabular}
\begin{tabular}{lcccccccccc}
Transition					&	Nature	& \multicolumn{5}{c}{Excitation energies} 						&	TBE\footnotemark[1]		& \multicolumn{2}{c}{Error wrt TBE}		&	Exp.\footnotemark[2]		\\
							\cline{3-7}														\cline{9-10}										
	 						&			&	exFCI	&	\sCI{4}		&	\sCI{5}		&	\sCI{6}		&	exDMC 		&			&	exFCI 	&	exDMC 	&	 \\
\hline
\ex{1}{A}{2}{}{n}{\pis}		&	Val.	&	$3.99$	&	$4.19(1)$	&	$4.07(2)$	&	$4.04(3)$	&	$4.02(3)$	&	$3.97$	&	$+0.02$	&	$+0.05(3)$	&	$4.07$	\\
\ex{1}{B}{2}{}{n}{3s}		&	Ryd.	&	$7.11$	&	$7.47(1)$	&	$7.38(2)$	&	$7.40(3)$	&	$7.30(3)$	&	$7.28$	&	$-0.17$	&	$+0.02(3)$	&	$7.11$	\\
\ex{1}{B}{2}{}{n}{3p}		&	Ryd.	&	$8.04$	&	$8.36(1)$	&	$8.28(2)$	&	$8.28(3)$	&	$8.21(3)$	&	$8.12$	&	$-0.08$	&	$+0.09(3)$	&	$7.97$	\\
\ex{1}{A}{1}{}{n}{3p}		&	Ryd.	&	$8.12$	&	$8.45(1)$	&	$8.31(2)$	&	$8.32(2)$	&	$8.24(2)$	&	$8.25$	&	$-0.13$	&	$-0.01(2)$	&	$8.14$	\\
\ex{1}{A}{2}{}{n}{3p}		&	Ryd.	&	$8.65$	&	$8.94(1)$	&	$8.90(2)$	&	$8.93(3)$	&	$8.66(3)$	&	$8.64$	&	$+0.01$	&	$+0.02(3)$	&	$8.37$	\\		
\hline
\ex{3}{A}{2}{}{n}{\pis}		&	Val.	&	$3.58$	&	$3.61(1)$	&	$3.63(2)$	&	$3.63(3)$	&	$3.65(2)$	&	$3.58$	&	$+0.00$	&	$+0.07(2)$	&	$3.50$	\\
\ex{3}{A}{1}{}{\pi}{\pis}	&	Val.	&	$6.10$	&	$6.15(1)$	&	$6.12(2)$	&	$6.13(3)$	&	$6.11(2)$	&	$6.07$	&	$+0.03$	&	$+0.04(2)$	&	$5.86$	\\
\ex{3}{B}{2}{}{n}{3s}		&	Ryd.	&	$6.95$	&	$7.30(1)$	&	$7.21(2)$	&	$7.20(3)$	&	$7.16(2)$	&	$7.12$	&	$-0.17$	&	$+0.04(2)$	&	$6.83$	\\
\ex{3}{B}{2}{}{n}{3p}		&	Ryd.	&	$7.87$	&	$8.18(1)$	&	$8.10(2)$	&	$8.09(3)$	&	$8.05(2)$	&	$7.98$	&	$-0.11$	&	$+0.07(2)$	&	$7.79$	\\
\ex{3}{A}{1}{}{n}{3p}		&	Ryd.	&	$8.01$	&	$8.33(1)$	&	$8.25(2)$	&	$8.23(3)$	&	$8.20(2)$	&	$8.13$	&	$-0.12$	&	$+0.07(2)$	&	$7.96$	\\
\end{tabular}
\end{ruledtabular}
\footnotetext[1]{Theoretical best estimates of Ref.~\onlinecite{Loos_2018} obtained from exFCI/AVTZ data corrected with the difference between CC3/AVTZ and CC3/d-aug-cc-pVQZ values.}
\footnotetext[2]{Various experimental sources, summarized in Ref.~\citenum{Robin_1985}.}
\end{table*}
%%% %%% %%%

%%% FIG 3 %%
\begin{figure*}
	\includegraphics[width=\linewidth]{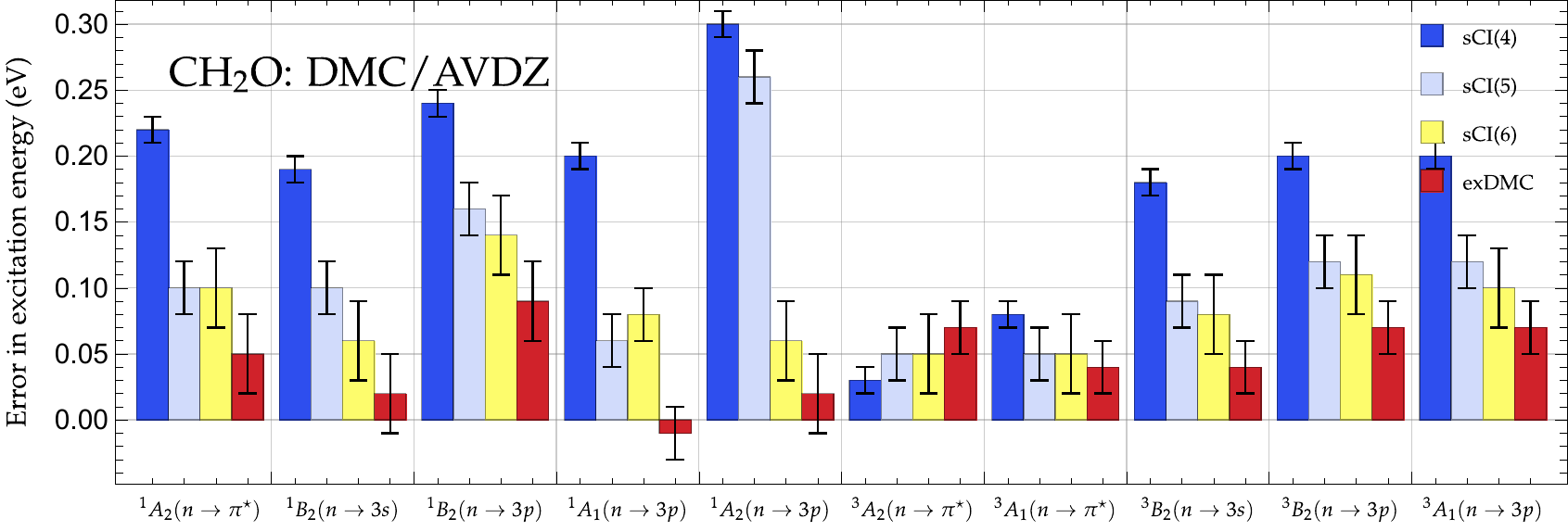}
	\caption{
	\label{fig:FigCH2O_1}
	Error (in eV) compared to the TBE of Ref.~\onlinecite{Loos_2018} for the five lowest singlet and five lowest triplet DMC vertical excitation energies of the formaldehyde molecule computed with the AVDZ basis set and various trial wave functions (see Table \ref{tab:PsiT}). 
	The error bar corresponds to one standard error.
	}
\end{figure*}
%%% %%% %%%

In order to test our QMC protocol on a larger molecule and check the validity of the conclusion drawn in the previous section, we have considered the formaldehyde molecule.
Similarly to water, formaldehyde is a very popular test molecule, \cite{Foresman_1992b, Hadad_1993, Head-Gordon_1994, Head-Gordon_1995, Gwaltney_1995, Wiberg_1998, Wiberg_2002, Peach_2008, Schreiber_2008,Shen_2009b, Caricato_2010, Li_2011, Leang_2012, Hoyer_2016} and stands as the prototype carbonyl dye with a low-lying $n \rightarrow \pis$ valence transition, well-separated from higher-lying excited states.
Moreover, formaldehyde has been previously studied at the FN-DMC level by Schautz et al. \cite{Schautz_2004e} 

Our all-electron DMC results for the formaldehyde molecule are reported in Table \ref{tab:CH2O-DMC} and represented in Fig.~\ref{fig:FigCH2O_1}.
For this molecule, we have studied a large number of singlet and triplet excited states with either valence or Rydberg characters (see Table \ref{tab:CH2O-DMC}).
However, we have restricted ourselves to the relatively compact AVDZ basis as we have previously observed that basis set effects are small within our QMC protocol (see Sec.~\ref{sec:water}).

For the lowest $n \rightarrow \pis$ singlet transition, the exDMC result yields a value of \IneV{$4.02(3)$}, which has to be compared to the TBE of {\IneV{$3.97$}, \cite{Loos_2018} and the MR-AQCC value of \IneV{$3.98$}. \cite{Muller_2001} 
However, it is significantly below the previous DMC estimate of \IneV{$4.24(2)$} reported by Schautz et al. \cite{Schautz_2004e} 
The latter discrepancy could be partially due to the use of both different structures and pseudopotentials within their DMC calculations.
%Interestingly, we obtain a similar value of \IneV{$4.19(1)$} with our smallest trial wave function \sCI{4} containing only roughly $23\,000$ determinants.
%This suggests that the discrepancy between the two DMC estimates might be directly related to the contrasted nodal quality of the two trial wave functions.

As shown in Fig.~\ref{fig:FigCH2O_1}, the dependency of the excitation energies with respect to $\PsiT$ is slightly more pronounced for \ce{CH2O} than for \ce{H2O}.
With the smallest trial wave function \sCI{4}, FN-DMC produces transition energies with errors as high as \IneV{$0.3$}, while the errors drop below \IneV{$0.1$} for all transitions when one considers the exDMC results.
Compared to the exFCI results reported in Table \ref{tab:CH2O-DMC}, we observe that FN-DMC provides, in most cases, a significant improvement.
Note however that, while exFCI usually underestimates the excitation energies, exDMC has a clear tendency to overcorrect them, hence yielding blue-shifted excitation energies.
Moreover, we emphasize that, for all the valence excitations, the exFCI error is positive while it is of opposite sign for the Rydberg states (similarly to water).
In exDMC, all the errors are positive (within statistical error), which means that the fixed-node error is always greater in the excited states than in the ground state.

As a concluding remark, we would like to mention that FN-DMC seems to perform similarly for valence and Rydberg excitations.
For the two lowest $n \rightarrow \pis$ triplet excitations with a clear valence nature, the trial wave function sensitivity looks less pronounced than for the other (Rydberg) excitations, as shown in Fig.~\ref{fig:FigCH2O_1}.
This could suggest that, for valence transitions (known to have a strong single-excitation character \cite{Loos_2018}), one could possibly rely on small-size $\PsiT$ in order to get accurate transition energies, even for larger molecules.
This latter point deserves further investigations.

%----------------------------------------------------------------
\section{
Conclusion
\label{sec:ccl}
}
%----------------------------------------------------------------
We have shown that, using a Jastrow-free QMC protocol relying on a deterministic and automatic construction of nodal surfaces, one can obtain accurate vertical transition energies within fixed-node DMC for small organic molecules.
We have illustrated our methodology on various singlet and triplet vertical transition energies of water and formaldehyde.
Our results for these two molecules evidence that accurate excitation energies can be obtained with relatively compact trial wave functions built with relatively small one-electron basis sets, thanks to a large cancelation of the fixed-node errors between ground and excited states.
Moreover, because the present protocol relies on sCI calculations to provide trial wave functions, the present approach has the indisputable advantage of being completely automatic and reproducible as one can eschew the non-linear stochastic optimization of the trial wave function which requires special care, especially for large systems.
Following the same methodology, we are currently investigating larger systems, and in particular simple cyanine dyes which are known to be particularly challenging for excited-state methods. \cite{LeGuennic_2015}

%----------------------------------------------------------------
\section*{
Supplementary material
\label{sec:SI}
}
%----------------------------------------------------------------
See {\SI} for optimized geometries, extrapolated total energies, and graphs of the DMC and CBS extrapolations.

%----------------------------------------------------------------
\begin{acknowledgments}
This work was performed using HPC resources from CALMIP (Toulouse) under allocations 2018-0510 and 2018-18005, and from GENCI-TGCC (Grant 2016-08s015).
AB was supported by the U.S.~Department of Energy, Office of Science, Basic Energy Sciences, Materials Sciences and Engineering Division, as part of the Computational Materials Sciences Program and Center for Predictive Simulation of Functional Materials.
\end{acknowledgments}
%----------------------------------------------------------------


\end{document}